\documentclass[prd,preprint,superscriptaddress,
  tightenlines,nofootinbib, eqsecnum]{revtex4-2}

\usepackage{amsmath}
\usepackage{amsfonts}
\usepackage{amssymb}
\usepackage{bm}
\usepackage{hyperref}
\usepackage{mathrsfs}
\usepackage{graphicx}
\usepackage{empheq}

\usepackage{ulem}
\usepackage[usenames]{color}

\usepackage[utf8]{inputenc}

\definecolor{darkgreen}{rgb}{0,0.5,0}

\hypersetup{
 bookmarks=true,         
 unicode=false,          
 pdftoolbar=true,        
 pdfmenubar=true,        
 pdffitwindow=false,     
 pdfstartview={FitH},    
 pdftitle={My title},    
 pdfauthor={Author},     
 pdfsubject={Subject},   
 pdfcreator={Creator},   
 pdfproducer={Producer}, 
 pdfkeywords={keyword1} {key2} {key3}, 
 pdfnewwindow=true,      
 colorlinks=true,       
 linkcolor=red,          
 citecolor=cyan,        
 filecolor=magenta,      
 urlcolor=darkgreen,           
 linktocpage=true
}

\newcommand{\PF}{\mathop{\mathrm{PF}}_{B=0}}

\newcommand{\intR}{\int_{r > \mathcal{R}}\!\!\frac{\dd^d\mathbf{x}}{\ell_0^\varepsilon} \,}
        
\allowdisplaybreaks

\DeclareSymbolFontAlphabet{\mathrsfs}{rsfs}
\DeclareMathAlphabet{\mathcal}{OMS}{cmsy}{m}{n}

\newcommand{\dd}{\mathrm{d}}
 
\newcommand{\de}{\mathrm{e}}

\begin{document}

\title{The Quadrupole Moment of Compact Binaries \\to the Fourth post-Newtonian Order\\I. Non-Locality in Time and Infra-Red Divergencies}

\author{Fran\c{c}ois \textsc{Larrouturou}}\email{francois.larrouturou@iap.fr}
\affiliation{$\mathcal{G}\mathbb{R}\varepsilon{\mathbb{C}}\mathcal{O}$, Institut d'Astrophysique de Paris, UMR 7095, CNRS, Sorbonne Universit{\'e}, 98\textsuperscript{bis} boulevard Arago, 75014 Paris, France}
\affiliation{Deutsches Elektronen-Synchrotron DESY, Notkestr. 85, 22607 Hamburg, Germany}

\author{Quentin \textsc{Henry}}\email{henry@iap.fr}
\affiliation{$\mathcal{G}\mathbb{R}\varepsilon{\mathbb{C}}\mathcal{O}$, Institut d'Astrophysique de Paris, UMR 7095, CNRS, Sorbonne Universit{\'e}, 98\textsuperscript{bis} boulevard Arago, 75014 Paris, France}
\affiliation{Max-Planck-Institute for Gravitational Physics (Albert-Einstein-Institute),
Am M\"{u}hlenberg 1, 14476 Potsdam, Germany}

\author{Luc \textsc{Blanchet}}\email{luc.blanchet@iap.fr}
\affiliation{$\mathcal{G}\mathbb{R}\varepsilon{\mathbb{C}}\mathcal{O}$, Institut d'Astrophysique de Paris, UMR 7095, CNRS, Sorbonne Universit{\'e}, 98\textsuperscript{bis} boulevard Arago, 75014 Paris, France}

\author{Guillaume \textsc{Faye}}\email{faye@iap.fr}
\affiliation{$\mathcal{G}\mathbb{R}\varepsilon{\mathbb{C}}\mathcal{O}$, Institut d'Astrophysique de Paris, UMR 7095, CNRS, Sorbonne Universit{\'e}, 98\textsuperscript{bis} boulevard Arago, 75014 Paris, France}

\date{\today}

\begin{abstract} 
With the aim of providing high accuracy post-Newtonian (PN) templates for the analysis of gravitational waves generated by compact binary systems, we complete the analytical derivation of the \textit{source type} mass quadrupole moment of compact binaries (without spins) at the fourth PN order of general relativity. Similarly to the case of the conservative 4PN equations of motion, we show that the quadrupole moment at that order contains a non-local (in time) contribution, arising from the tail-transported interaction entering the conservative part of the dynamics. Furthermore, we investigate the infra-red (IR) divergences of the quadrupole moment. In a previous work, this moment has been computed using a Hadamard partie finie procedure for the IR divergences, but the knowledge of the conservative equations of motion indicates that those divergences have to be dealt with by means of dimensional regularization. This work thus derives the difference between the two regularization schemes, which has to be added on top of the previous result. We show that unphysical IR poles start to appear at the 3PN order, and we determine all of these up to the 4PN order. In particular, the non-local tail term comes in along with a specific pole at the 4PN order. It will be proven in a companion paper that the poles in the source-type quadrupole are cancelled in the physical \textit{radiative type} quadrupole moment measured at future null infinity.
\end{abstract}


\maketitle

\section{Introduction}\label{sec:introduction}

The theory of gravitational waves (GW) generated by binary systems of compact objects has been developed using perturbative methods in classical general relativity (see~\cite{Maggiore, BlanchetLR, BuonSathya15,Porto16} for reviews). In particular the post-Newtonian (PN) approximation is a major and widely developed technique for computing analytically the dynamics and GW emission of compact binaries. The state of the art on the conservative dynamics are the equations of motion of point-particle binaries at the 4PN (fourth-post-Newtonian) order~\cite{DJS14, DJS16, BBBFMa, BBBFMc, MBBF17, GLPR16, PR17,FS19, FPRS19,Blumlein20}. Recently progresses have been made so that the equations of motion are now determined up to two unknown parameters at 5PN order, and up to six unknowns at 6PN order~\cite{Blumlein21, bini2020sixth}. Currently the field is also evolving thanks to new methods coming from effective field theory and scattering amplitudes, naturally combined with the classical post-Minkowskian approximation~\cite{Bern:2019nnu}. 

This paper is concerned with the GW emission aspect, which is directly related to the data analysis of GW detectors. Here, the state of the art is the 3PN approximation in the waveform, \textit{beyond} the Einstein quadrupole formula~\cite{BDI95, WW96, B96, BI04mult, BDEI05dr}. Actually, the flux and orbital phase evolution due to gravitational radiation are known to the 3.5PN order~\cite{B98tail, BFIJ02}, as well as the dominant mass-quadrupole mode $(\ell, m) = (2,2)$~\cite{BFIS08, FMBI12}, the current-quadrupole mode $(2,1)$~\cite{HFB_courant} and the mass-octupole ones $(3,3)$ and $(3,1)$~\cite{FBI15}.
The gravitational flux and mass quadrupole were confirmed at the 2PN order by means of effective field theory techniques~\cite{LMRY19}.

Extending the GW emission up to the 4PN (and even 4.5PN~\cite{MBF16}) order is the target of the present program. A central part of this program is of course the control of the mass-type quadrupole moment of the system with the 4PN precision. This computation faces subtle issues regarding the choice and proper use of regularization schemes, both for the ultra-violet (UV) and infra-red (IR) divergences. Recently, a preliminary calculation of the source mass-type quadrupole moment of compact binaries (of spinless bodies) at the 4PN order has been tackled~\cite{MHLMFB20}. In this calculation, the UV divergences, appearing because of the point-like structure of the source (modelling compact objects with negligible internal structure), were properly treated by means of the powerful dimensional regularization. However, the IR ones were regularized with the Hadamard partie finie (PF) regularization procedure~\cite{Hadamard}. Furthermore, in the calculation of Ref.~\cite{MHLMFB20}, the non-local-in-time contributions, due to retarded correlations over arbitrarily large time spans in  the dynamics of the source~\cite{BD88}, were neglected. 
 
In the present paper and the next one~\cite{DDR_radiatif}, we complete the derivation of the 4PN source mass-quadrupole moment of compact binaries. More precisely, our goals are two-fold: 
\begin{enumerate}
	\item To derive the non-local (in time) effect in the source mass quadrupole moment which is due to the radiation modes associated with propagating tails at infinity. This effect is the analogue of the one occurring in the conservative equations of motion and the Lagrangian/Hamiltonian at the 4PN order~\cite{DJS14, GLPR16, BBBFMa};
	\item To compute all the contributions to the quadrupole moment due to the IR divergences, implementing a dimensional regularization scheme rather than the Hadamard PF scheme adopted in~\cite{MHLMFB20}. Such procedure leads to the appearance of specific IR poles which start to arise at the 3PN order and play a crucial role at the 4PN order.
\end{enumerate}
In the follow-up paper~\cite{DDR_radiatif}, we investigate the fate of these IR poles in the radiative-type quadrupole moment, which represents the actual observable moment at future null infinity, up to the 4PN order. We will find that they are exactly canceled by radiation contributions due to non-linear propagation effects (the most important ones being coined as ``tails-of-tails'' and ``tails-of-memory'') and that we can therefore safely define a three-dimensional ``\textit{renormalized}'' mass quadrupole moment at the 4PN order, which will constitute a basic ingredient in the construction of 4PN GW templates.

The central object investigated in this work is thus the \textit{source-type} mass-quadrupole moment, defined in $d$ spatial dimensions by the expression~\cite{BDEI04, MHLMFB20, HFB_courant}\footnote{The overbar denotes the formal PN expansion; superscript parenthesis $(n)$ denote time derivatives; the hat refers to the symmetric-trace-free (STF) product, \textit{e.g.} $\hat{x}_{ija} \equiv \text{STF}[x_i x_j x_a]$; $\text{PF}_{B=0}$ denotes the Hadamard partie finie with regulator $(r/r_0)^B$ and associated length scale $r_0$ ($B\in \mathbb{C}$); the characteristic dimensional regularization length scale is $\ell_0$.}
\begin{align}\label{eq:Iij_d_dim}
I_{ij} &= 
\frac{d-1}{2(d-2)}\PF\int\frac{\dd^d \mathbf{x}}{\ell_0^{d-3}}\left(\frac{r}{r_0}\right)^B \biggr\{
\hat{x}_{ij}\,\overline{ \Sigma}_{[2]} 
-\frac{4(d+2)} {c^2\, d(d+4)}\,\hat{x}_{ija}\, \overline{\Sigma}^{(1)}_{a[3]} \nonumber\\
&\qquad\qquad
+\frac{2(d+2)} {c^4\,d(d+1)(d+6)} \,\hat{x}_{ijab}\, \overline{\Sigma}^{(2)}_{ab[4]} - \frac{4B(d-3)(d+2)}{c^2\,(d-1)d(d+4)}\,\hat{x}_{ija}\,\frac{x_b}{r^2} \,\overline{\Sigma}_{ab[3]} \biggr\}\,.
\end{align}
We let the reader refer to~\cite{MHLMFB20} for a comprehensive review of the definitions and properties of the quantities entering the source quadrupole moment, as well as its computation using the IR Hadamard and UV dimensional regularizations. Let us just emphasize a few points.

The main quantity over which the source quadrupole integrates is the pseudo stress-energy tensor $\tau^{\mu\nu}$ in harmonic coordinates composed of a matter and a gravitational part,
\begin{equation}\label{eq:tau}
\tau^{\mu\nu} \equiv \vert g\vert T^{\mu\nu}+\frac{c^4}{16\pi
	G}\,\Lambda^{\mu\nu}\,,
\end{equation}
where $T^{\mu\nu}$ is the matter stress-energy tensor and $\Lambda^{\mu\nu}$ the non-linear gravitational source term of the Einstein field equations. It enters Eq.~\eqref{eq:Iij_d_dim} through the following PN-expanded source densities:
\begin{equation}\label{eq:Sigmad}
\overline{\Sigma} = \frac{2}{d-1}\frac{(d-2)\overline{\tau}^{00}+\overline{\tau}^{ii}}{c^2}\,,\qquad\overline{\Sigma}_i = \frac{\overline{\tau}^{0i}}{c}\,, \qquad\overline{\Sigma}_{ij} = \overline{\tau}^{ij}\,.
\end{equation}
The PN expansions implicit in~\eqref{eq:Iij_d_dim} read as (with $\Gamma$ the Euler function)
\begin{equation}\label{eq:PNseries}
\overline{\Sigma}_{[\ell]}(\mathbf{x},t)
\equiv \sum_{k=0}^{+\infty}\frac{1}{2^{2k} k!}\frac{\Gamma(\ell+\frac{d}{2})}{\Gamma(\ell+\frac{d}{2}+k)}
\left(\frac{r}{c}\frac{\partial}{\partial
	t}\right)^{2k}\overline\Sigma(\mathbf{x},t)\,.
\end{equation}

The partie finie procedure $\text{PF}_{B=0}$ comes from the matching between near zone and exterior zone. It is crucial for the proper definition of the multipole moments in 3 dimensions~\cite{B98mult}. The derivation of the equations of motion~\cite{BBBFMc, MBBF17} indicated that the PF operator should be kept even in $d$ dimensions. It was also shown there that, in $d$ dimensions, the limit $B\to 0$ is finite in physical quantities (without pole $1/B$); so, in the end the whole procedure is equivalent to the usual dimensional regularization. We had however to keep the PF operator as explicit in the expression of the quadrupole moment~\eqref{eq:Iij_d_dim}. Here, similarly to the equations of motion~\cite{BBBFMc, MBBF17}, we apply first the PF process when $B\to 0$ on the $d$ dimensional expression and, second, the usual dimensional regularization when $\varepsilon\equiv d-3 \to 0$. We call this mixed regularization the ``$B\varepsilon$'' regularization. Note that the last term in~\eqref{eq:Iij_d_dim}, proportional to both $B$ and $d-3$, will be shown to play no role with this particular regularization. 

This paper is organized as follows. In the next Section~\ref{sec:nonlocal}, we derive the non-local in time part of the source quadrupole, relying on results derived in~\cite{BBBFMc}. We then perform the proper IR dimensional regularization for all the various categories of terms composing the quadrupole in Section~\ref{sec:sourcequad}. The final result is presented in Section~\ref{sec:final} in the form of a pole followed by a finite part contribution, which will be the starting point for the final renormalization of the quadrupole in the next paper~\cite{DDR_radiatif}. Appendix~\ref{app:riesz} present technical formulas generalizing the Riesz formula in $d$ dimensions. Appendix~\ref{app:IR_shift} gives the expression of the local part of the IR shift coming from the 4PN equations of motion.

\section{Non-locality in time of the source quadrupole moment}
\label{sec:nonlocal}

Crucial to the completion of the ambiguity-free equations of motion at the 4PN order was the proper inclusion of the tail effect, \textit{i.e.} the non-local in time back-scattering of emitted gravitational waves, modifying the conservative dynamics of the system at the current time~\cite{BBBFMc}. This effect enters at the 4PN order in the near-zone metric and, as such, plays a key role in the computation of the 4PN source mass quadrupole by introducing a non-local term, together with a pole.

\subsection{The tail effect in the conservative 4PN equations of motion}

Let us first recall how the tail effect is included in the near-zone metric and Fokker action for the conservative dynamics. In Ref.~\cite{BBBFMc}, it was found that the PN-expanded gravitational field in harmonic coordinates ($h^{\mu\nu}\equiv\sqrt{-g}g^{\mu\nu}-\eta^{\mu\nu}$, which is such that $\partial_\nu h^{\mu\nu}=0$) contains the following pieces responsible for tails at the 4PN order:
\begin{subequations}\label{eq:h2harmtail}
	\begin{align}
\overline{h}^{00ii}_\text{tail} &= \frac{8 G^2M}{15 c^{10}} \, x^{ij}
		\int_0^{+\infty}\dd\tau\left[ L_\varepsilon(\tau) +
		\frac{61}{60}\right]I^{(7)}_{ij}(t-\tau) +
		\mathcal{O}\left(\frac{1}{c^{12}}\right) \,,\\
\overline{h}^{0i}_\text{tail} &= - \frac{8 G^2M}{3 c^{9}} \,x^j
		\int_0^{+\infty}\dd\tau\left[ L_\varepsilon(\tau) +
		\frac{107}{120}\right]I^{(6)}_{ij}(t-\tau) +
		\mathcal{O}\left(\frac{1}{c^{11}}\right) \,,\\
\overline{h}^{ij}_\text{tail} &= \frac{8 G^2M}{c^{8}}
		\int_0^{+\infty}\dd\tau\left[ L_\varepsilon(\tau) +
		\frac{4}{5}\right]I^{(5)}_{ij}(t-\tau) +
		\mathcal{O}\left(\frac{1}{c^{10}}\right) \,,\label{eq:h2harmtailij}
	\end{align}
\end{subequations}
where we have posed $h^{00ii} \equiv \frac{2}{d-1}[(d-2)h^{00} + h^{ii}]$ and introduced the short-hand notation for the logarithmic divergence with the associated pole $\propto \varepsilon^{-1}\equiv (d-3)^{-1}$:
\begin{equation}\label{eq:notL}
	L_\varepsilon(\tau) \equiv \ln\left(\frac{c\sqrt{\bar{q}}\,\tau}{2\ell_0}
	\right) - \frac{1}{2\varepsilon}\,,
\end{equation}
where $\bar{q} \equiv 4\pi\,\de^{\gamma_\text{E}}$ with $\gamma_\text{E}$ the Euler constant, vanishingly small terms $\mathcal{O}(\varepsilon)$ being neglected. 

In our set-up and notation, the two and only two series of multipole moments describing the radiation field generated by an isolated source are the so-called canonical moments, $M_L$ and $S_L$. Those differ from the source moments $I_L$ and $J_L$, \textit{e.g.}~\eqref{eq:Iij_d_dim} and, thus, in principle, the quadrupole moment in~\eqref{eq:h2harmtail} should rather be viewed as the canonical moment and denoted $M_{ij}$. However, we take here advantage that $M_{ij}$ and the source moment $I_{ij}$ are equivalent at Newtonian order (and even up to 2PN order).

Still in Ref.~\cite{BBBFMc}, we next applied a gauge transformation, at quadratic order, for the particular interaction $M \times I_{ij}$, so designed as to transfer all relevant tail terms in the ``$00ii$'' component of the metric. Namely, we posed ${h'}^{\mu\nu} = h^{\mu\nu} + \partial^\mu\epsilon_\text{tail}^\nu + \partial^\nu\epsilon_\text{tail}^\mu - \eta^{\mu\nu} \partial_\rho\epsilon_\text{tail}^\rho$, with PN-expanded gauge vector given by the following tail pieces (used in~\cite{BBBFMc} but published here for the first time):
\begin{subequations}\label{eq:epsilontail}
	\begin{align}
	\overline{\epsilon}^{0}_\text{tail} &= - \frac{2G^2M}{3c^{9}} \,x^{ij}
	\int_0^{+\infty}\dd\tau\left[ L_\varepsilon(\tau) +
	\frac{37}{60}\right]I^{(6)}_{ij}(t-\tau) +
	\mathcal{O}\left(\frac{1}{c^{11}}\right) \,,\\
	\overline{\epsilon}^{i}_\text{tail} &= - \frac{4 G^2M}{c^{8}} \,x^j
	\int_0^{+\infty}\dd\tau\left[ L_\varepsilon(\tau) +
	\frac{4}{5}\right]I^{(5)}_{ij}(t-\tau) +
	\mathcal{O}\left(\frac{1}{c^{10}}\right)\,.
	\end{align}
\end{subequations}
In the new gauge, the 4PN tail effect is thus entirely described by the single scalar potential ${h'}^{00ii}_\text{tail}$ (or, equivalently, by the $00$ component of the covariant metric ${g'}_{00}^\text{tail}$), which becomes
\begin{align}
	{\overline{h}'}^{00ii}_\text{tail} &= \frac{16 G^2M}{5 c^{10}} \,x^{ij}
	\int_0^{+\infty}\dd\tau\left[ L_\varepsilon(\tau) +
	\frac{41}{60}\right]I^{(7)}_{ij}(t-\tau) +
	\mathcal{O}\left(\frac{1}{c^{12}}\right) \,.
\end{align}
This tail piece in the metric yields the tail term in the conservative Fokker action, which is found to be manifestly symmetric under time reversal, 
\begin{align}\label{eq:Stail}
	S_\text{tail} &= \frac{G^2 M}{5 c^8} \int_{-\infty}^{+\infty}\dd t
	\,I_{ij}^{(3)}(t) \int_0^{+\infty} \dd\tau \biggl[
	L_\varepsilon(\tau) + \frac{41}{60} \biggr] \left(I_{ij}^{(4)}(t-\tau) - I_{ij}^{(4)}(t+\tau)\right) \,.
\end{align}
An elegant alternative form is provided by the Hadamard partie finie (Pf) integral 
\begin{align}\label{eq:Stail_alt}
S_\text{tail} = \frac{G^2 M}{5 c^8} \,\mathop{\text{Pf}}_{\tau_0}
\int\!\!\!\int \frac{\dd t\dd t'}{\vert t-t'\vert}
I_{ij}^{(3)}(t)\,I_{ij}^{(3)}(t')\,,
\end{align}
with the Hadamard regularization scale $\tau_0 = \frac{2\ell_0}{c \sqrt{\bar{q}}}\,\text{exp}[\frac{1}{2\varepsilon}-\frac{41}{60}]$.

\subsection{Direct tail term in the 4PN mass quadrupole moment}

Inserting Eqs.~\eqref{eq:h2harmtail} into the quadrupole moment~\eqref{eq:Iij_d_dim}, it is straightforward to see that the only 4PN effect comes from the first term, namely
\begin{align}\label{eq:Iijtail0}
I_{ij}^\text{tail} = \frac{d-1}{2(d-2)} \mathop{\mathrm{PF}}_{B=0}\int \dd^d \mathbf{x} \,\left(\frac{r}{r_0}\right)^B \, \hat{x}_{ij}\,\overline{\Sigma}_\text{tail} + \cdots\,,
\end{align}
where only the Newtonian term in the PN series~\eqref{eq:PNseries} needs to be considered, and where the ellipsis denote other terms that do not participate to the effect at 4PN order. Considering the order of appearance of the tail integrals in the metric, the only terms in the non-linear source $\Lambda^{\mu\nu}$ which may contribute to the effect at the 4PN order are $\Lambda^{00} = - h^{ab}\partial_{ab} h^{00}+ \frac{1}{4} \partial_a h^{00} \partial_a h^{bb}$ and $\Lambda^{ii}= -\frac{1}{4} (d-2) \partial_a h^{00} \partial_a h^{bb}$. Now, only the components of the metric that are proportional to $1/c^8$, namely $h_\text{tail}^{ab}$, can subsist at this accuracy level. They are directly given by~\eqref{eq:h2harmtailij} and are obviously trace-free. The tail sector of the effective non-linear source in~\eqref{eq:Iijtail0} thus reduces to
\begin{equation}
\overline{\Sigma}_\text{tail} = -\frac{c^2}{16 \pi G} \frac{2(d-2)}{d-1}\, h_\text{tail}^{ab}\partial_{ab} h^{00} + \cdots\,.
\end{equation}
On the other hand, to the lowest order, we have $h^{00} = - \frac{2(d-1)}{d-2} \frac{V}{c^2} + \cdots$, with $V$ being, at leading order, the Newtonian potential in $d$ dimensions. Using also the fact that $h_\text{tail}^{ab}$ is a function of time only, we find that the 4PN tail contribution in the quadrupole moment reads
\begin{align}\label{eq:intIijtail}
I_{ij}^\text{tail} &= \frac{d-1}{8\pi G(d-2)} \,h^{ab}_\text{tail} \PF \int \dd^d \mathbf{x} \left(\frac{r}{r_0}\right)^B \hat{x}^{ij}\partial_{ab} V\,.
\end{align}
Injecting the Newtonian potential $V = \tilde{k} \,G\,m_1 \frac{2(d-2)}{d-1}\,r_1^{2-d} + 1 \leftrightarrow 2$, with $\tilde{k}=\pi^{1-\frac{d}{2}} \Gamma(\frac{d}{2}-1)$, the term~\eqref{eq:intIijtail} can be computed by means of the ``generalized Riesz integrals'' presented in Appendix~\ref{app:riesz}, more precisely Eq.~\eqref{rieszgeneral} with $\boldsymbol{y}_2=0$. Finally, applying the operator PF at $B=0$ which reduces here to a simple limit when $B\to 0$,\footnote{The calculation boils down to the single simple elementary integral
  $$\frac{\tilde{k}}{2\pi} \PF \int \dd^d \mathbf{x} \left(\frac{r}{r_0}\right)^B  \hat{x}^{ij}\, r_1^{2-d} =- \frac{1}{d+4} \,\bm{y}_1^2 \,y_1^{\langle ij\rangle}\,.$$}
   we find at 4PN order
\begin{align}\label{eq:Itaildirect}
I_{ij}^\text{tail} =  - \frac{8 G^2M}{c^{8}} \,\int_0^{+\infty}\dd\tau\left[ L_\varepsilon(\tau) +
\frac{4}{5}\right]\left( \frac{4}{d+4} I_{k\langle i}(t) I^{(5)}_{j\rangle k}(t-\tau) + \frac{I(t)}{d} I^{(5)}_{ij}(t-\tau)\right)\,.
\end{align}
Here, the STF quadrupole moment reads $I_{ij} = m_1 \,y_1^{\langle ij\rangle} + 1\leftrightarrow 2 + \mathcal{O}(c^{-2})$; we have also introduced the Newtonian moment of inertia $I = m_1 \,\bm{y}_1^2 + 1\leftrightarrow 2 + \mathcal{O}(c^{-2})$.

\subsection{Indirect tail contribution due to a 4PN shift and total tail effect}

Now, an important point for our purpose is to remark that the gauge transformation that has been applied to the 4PN equations of motion, with gauge vector~\eqref{eq:epsilontail}, will induce a spatial shift of the particles' world-lines, which has to be taken into account when evaluating the quadrupole moment. Applying such a shift is crucial to ensure the coherence between the coordinate systems used in the derivation of the equations of motion and the multipole moments. Denoting the value on particle 1 of the gauge vector as $\epsilon_\text{tail\,1}^\mu \equiv \epsilon_\text{tail}^\mu[t,\bm{y}_1(t)]$, the spatial shift $\boldsymbol{\zeta}_1(t)\equiv\boldsymbol{y}_1(t)-\boldsymbol{y}'_1(t)$ will be given in general by $\zeta_\text{1}^i = - \epsilon^i_\text{tail\,1} + v_1^i\,\epsilon^{0}_\text{tail\,1}/c + \mathcal{O}(\epsilon^2)$. It is clear from the $1/c$ factors of Eq.~\eqref{eq:epsilontail} that only the first term contributes to 4PN order; hence the shift reads
\begin{align}\label{eq:nlshift}
\overline{\zeta}_1^i = \frac{4 G^2M}{c^{8}} \,y_1^j
\int_0^{+\infty}\dd\tau\left[ L_\varepsilon(\tau) +
\frac{4}{5}\right]I^{(5)}_{ij}(t-\tau)  +
	\mathcal{O}\left(\frac{1}{c^{10}}\right)\,,
\end{align}
and the contribution brought about by this shift in the quadrupole moment at 4PN is
\begin{align}
\delta_\zeta I_{ij}\equiv 2 m_1 y_1^{\langle i} \overline{\zeta}^{\,j\rangle}_{1} + 1\leftrightarrow 2 +
\mathcal{O}\left(\frac{1}{c^{10}}\right)\,.
\end{align}
Therefore, again denoting by $I = m_1 \,\bm{y}_1^2 + 1\leftrightarrow 2 + \mathcal{O}(c^{-2})$ the Newtonian moment of inertia, we obtain at 4PN order
\begin{align}\label{eq:Itailshift}
\delta_\zeta I_{ij} = \frac{8 G^2M}{c^{8}} \,\int_0^{+\infty}\dd\tau\left[ L_\varepsilon(\tau) +
\frac{4}{5}\right]\left( I_{k\langle i}(t) I^{(5)}_{j\rangle k}(t-\tau) + \frac{I(t)}{d} I^{(5)}_{ij}(t-\tau)\right)\,.
\end{align}

As the computation carried out in~\cite{MHLMFB20} was purely local, none of the non-local effects (neither the direct effect of the tails nor the contribution of the non-local shift) were included in this preliminary result. Combining the two new contributions of this work~\eqref{eq:Itaildirect} and~\eqref{eq:Itailshift}, we thus have to add a non-local in time term to the mass quadrupole moment previously computed in~\cite{MHLMFB20} which reads explicitly at 4PN [neglecting terms $\mathcal{O}(\varepsilon)$]
\begin{align}\label{Itail}
I_{ij}^\text{non-loc} \equiv I_{ij}^\text{tail} + \delta_\zeta I_{ij} = \frac{24 G^2M}{7 c^{8}} \,I_{k\langle i}(t)\int_0^{+\infty}\dd\tau\left[ L_\varepsilon(\tau) +
\frac{74}{105}\right] I^{(5)}_{j\rangle k}(t-\tau)\,.
\end{align}
An interesting point is that the contribution of the moment of inertia $I(t)$ cancels out, but there remains a pole. The next section and the following paper~\cite{DDR_radiatif} will show how this pole finally combines with other poles coming from IR dimensional regularization to disappear from the observable radiative moment.

Note finally that $I_{ij}^\text{non-loc}$ contains both conservative and dissipative effects, together with a purely instantaneous (non-tail) piece. Those will be specified and computed on quasi-circular orbits in the companion paper~\cite{DDR_radiatif}.

\section{IR regularization of the source quadrupole moment}\label{sec:sourcequad}

The quadrupole moment in $d$ dimensions given by~\eqref{eq:Iij_d_dim}, when PN-expanded using Eqs.~\eqref{eq:PNseries}, and after some suitable integrations by part, can be decomposed into four different types of contributions (see App.~C of~\cite{MHLMFB20} for the exhaustive list of all those terms):
\begin{enumerate} 
\item Volume terms, where the integrands are made of products of derivatives of elementary potentials. The complete list of potentials is provided in Appendix~A of~\cite{MHLMFB20};\footnote{The derivatives of those products of potentials are treated in the context of the theory of distributions. They can thus include a distributional sector, made with Dirac distributions. Those are naturally regarded as compact terms.}
\item Compact-support terms, where the integrands are proportional to Dirac distributions modeling the compact objects, multiplying products of derivatives of elementary potentials;
\item Surface terms, where the integrands are total spatial derivatives and can be replaced by their expansions at spatial infinity;
\item The ``extra'' term, which is the last term of~\eqref{eq:Iij_d_dim}, proportional to $B(d-3)$, and formally zero in Hadamard's sense, in 3 dimensions within a UV dimensional regularization scheme.
\end{enumerate}
Since the calculations concern only the IR bound, we only need the appropriate accurate values for the potentials when expanded at spatial infinity. Interestingly, the integral giving the value of a potential at the location of particles extends up to infinity, thus this applies also to the case of compact terms, which may be affected by the change of regularization [see Eq.~\eqref{eq:Poisson_comp} below]. 

In Ref.~\cite{MHLMFB20}, the applied IR regularization scheme was a pure Partie Finie (PF) one as $B\rightarrow 0$ (with poles $1/B$ discarded) on the three-dimensional expression of~\eqref{eq:Iij_d_dim}. Instead, we resort here to the mixed ``$B\varepsilon$'' regularization which consists of computing first the limit $B\to 0$ on the $d$-dimensional expression of the quadrupole~\eqref{eq:Iij_d_dim}, and only then apply the dimensional regularization when $\varepsilon\to 0$, of course keeping track of all the poles $1/\varepsilon$. By contrast with the pure PF regularization, we expect from the work~\cite{BBBFMc, MBBF17} on equations of motion that the first limit $B\to 0$, \textit{i.e.}, performed on the top of dimensional regularization, will be finite in that case (no pole $1/B$), which is confirmed by our calculations below.

Each different type of terms is to be dimensionally regularized (following the $B\varepsilon$ scheme) with specific techniques, which are exposed in the rest of the section. In a nutshell, we aim at computing
\begin{equation}\label{eq:DIij_def}
\mathcal{D}I_{ij} \equiv I_{ij}^{B\varepsilon} -  I_{ij}^\text{Had} = \mathcal{D}I_{ij}^\text{Vol} + \mathcal{D}I_{ij}^\text{Comp} + \mathcal{D}I_{ij}^\text{Surf} + \mathcal{D}I_{ij}^\text{extra}\,,
\end{equation}
where $I_{ij}^{B\varepsilon}$ represents the source mass quadrupole properly regularized following the $B\varepsilon$ prescription, $I_{ij}^\text{Had}$ is an abuse of notation for the result of~\cite{MHLMFB20}, \textit{i.e.} the source quadrupole computed with dimensional regularization for the UV divergences and the Hadamard regularization for the IR ones. The terms in the right-hand-side (RHS) represent the differences for the four types of terms 1 to 4.

\subsection{Volume terms}

The most numerous terms to regularize are the volume terms. Their regularizations consist in two parts: (i) one has to compute the general expression of the difference between the two regularization schemes $B\varepsilon$ and Had for each term, and then (ii) inject the accurate $d$-dimensional values for the potentials expanded at spatial infinity.

\subsubsection{Dimensional regularization of volume terms}

Let us consider a generic volume term
\begin{equation}
\mathcal{V} = \int\!\frac{\dd^d\mathbf{x}}{\ell_0^\varepsilon} \left(\frac{r}{r_0}\right)^B F(\mathbf{x},t)\,.
\end{equation}
The function $F(\mathbf{x},t)$ is some product of (derivatives of) potentials and some $\hat{x}_L$; below, we will drop the time dependence as it plays no role in the regularization procedure. As we investigate the difference between IR regularization schemes, we can restrict the integral to $r > \mathcal{R}$, where $\mathcal{R}$ is an arbitrary constant scale, significantly larger than the distances of the particles to the origin, $\vert\bm{y}_A\vert$, so that we do not have to consider the problem coming from the point-particle approximation, already dealt with in~\cite{MHLMFB20}.

In Ref.~\cite{MHLMFB20}, we have computed the pure PF regularization
\begin{equation}
\mathcal{V}^\text{Had} = \PF\int_{r > \mathcal{R}}\!\!\dd^3\mathbf{x} \left(\frac{r}{r_0}\right)^B F^{(d=3)}(\mathbf{x})\,,
\end{equation}
where $ F^{(d=3)}$ denotes naturally the three-dimensional limit of $F$ obtained by performing the PN iteration in 3 dimensions and given by~\eqref{eq:F3}. On the other hand, in this work we consider the mixed regularization scheme $B\varepsilon$,
\begin{equation}
\mathcal{V}^{B\varepsilon} = \PF\intR \left(\frac{r}{r_0}\right)^B F(\mathbf{x})\,.
\end{equation}
Thus, for each volume term, we are to compute the difference of regularization schemes
\begin{equation}
\mathcal{D}\mathcal{V} \equiv \mathcal{V}^{B\varepsilon}-\mathcal{V}^\text{Had}\,.
\end{equation}
Since, in the limit $\varepsilon\to 0$, the complementary integrals over $r<\mathcal{R}$ agree with each other, this difference should be independent of the cut-off scale $\mathcal{R}$. We conclude that $\mathcal{D}\mathcal{V}$ is the proper quantity that we have to add to the volume terms computed in~\cite{MHLMFB20}.

As it appears at the 4PN order, the functions $F$ we consider admit generic far-zone (or multipolar) expansions in $d$ dimensions when $r \equiv \vert \mathbf{x}\vert \rightarrow +\infty$, namely\footnote{By which we really mean that, for any $N\in\mathbb{N}$, we can write $$F(\mathbf{x}) = \sum_{p=-p_0}^N\sum_{q=q_0}^{q_1}\frac{\ell_0^{q\varepsilon}}{r^{p+q\varepsilon}}\,\varphi^{(\varepsilon)}_{p,q}(\mathbf{n}) + o\left(\frac{1}{r^N}\right)\,,$$ where $p_0\in\mathbb{N}$ indicates the maximal order of the IR divergence, and $q_0$, $q_1\in\mathbb{Z}$ represent a finite range of values for $q$ depending implicitly on $N$.}
\begin{equation}
F(\mathbf{x}) = \sum_{p,q}\frac{\ell_0^{q\varepsilon}}{r^{p+q\varepsilon}}\,\varphi^{(\varepsilon)}_{p,q}(\mathbf{n})\,,
\end{equation}
with coefficients that can contain poles $\propto 1/\varepsilon$, \textit{i.e.} of the type
\begin{equation}
\varphi^{(\varepsilon)}_{p,q}(\mathbf{n}) = \frac{1}{\varepsilon}\,\psi^{(-1)}_{p,q}(\mathbf{n}) + \psi^{(\varepsilon)}_{p,q}(\mathbf{n})\,.
\end{equation}
We have verified that no double poles $\propto 1/\varepsilon^2$ appear at the 4PN order. The coefficients $\psi^{(-1)}_{p,q}$ of the pole are naturally defined with no dependence upon $\varepsilon$, while the $\psi^{(\varepsilon)}_{p,q}$ are finite when $\varepsilon\to 0$. An important point is that, despite the poles, the 3 dimensional limits of the functions $F$ are finite, as clear from their expressions given in the App.~C of~\cite{MHLMFB20} and explicitly verified in our computation. This means that (for all $p$)
\begin{subequations}\label{eq:relationdto3}
\begin{equation}
\sum_q \psi^{(-1)}_{p,q}(\mathbf{n}) = 0\,.
\end{equation} 
Furthermore, by posing
\begin{equation}
\sum_q \psi^{(\varepsilon=0)}_{p,q}(\mathbf{n}) \equiv  \varphi_p(\mathbf{n})\,,
\qquad
\sum_q q\,\psi^{(-1)}_{p,q}(\mathbf{n}) \equiv - \varphi^{\ln}_p(\mathbf{n})\,,
\end{equation} 
\end{subequations}
we get the corresponding logarithmic expansion in 3 dimensions 
\begin{equation}\label{eq:F3}
F^{(d=3)} = \sum_p \frac{1}{r^p}\left[\varphi_p(\mathbf{n}) + \varphi^{\ln}_p(\mathbf{n})\,\ln\left(\frac{r}{\ell_0}\right)\right]\,.
\end{equation} 
Note that, as a confirmation of the absence of double poles in the $d$-dimensional expressions, no squared logarithms appeared in the Hadamard computation.

Using the relations~\eqref{eq:relationdto3} linking the $d$ and $3$ dimensional quantities, we obtain
\begin{equation}\label{eq:DiffV}
\mathcal{D}\mathcal{V}  = 
\sum_{q\neq 1}\frac{1}{\varepsilon}\left[\frac{1}{q-1} - \varepsilon\ln\left(\frac{r_0}{\ell_0}\right)\right]\!\int\!\dd \Omega_{d-1}\,\varphi^{(\varepsilon)}_{3,q}(\mathbf{n})
- \frac{1}{2}\ln^2\left(\frac{r_0}{\ell_0}\right)\int\!\dd\Omega_2\,\varphi^{\ln}_3(\mathbf{n}) + \mathcal{O}\left(\varepsilon\right)\,,
\end{equation}
where $\dd \Omega_{d-1}$ denotes the $(d-1)$-dimensional solid angle element. As expected, we find that, modulo $\mathcal{O}(\varepsilon)$-terms, this difference does not depend on the cut-off scale $\mathcal{R}$. The result~\eqref{eq:DiffV} generalizes that of Eq.~(2.11) in~\cite{BBBFMc}, to include integrands containing poles. Note that the value $q=1$ has to be excluded from the sum in~\eqref{eq:DiffV}. This is \emph{not} an artificial requirement to ensure that the formula is well-defined: it comes naturally out of the derivation. A last use of~\eqref{eq:relationdto3} yields the more compact equivalent form
\begin{equation}\label{eq:DV_compact}
\mathcal{D}\mathcal{V} = \frac{1}{\varepsilon}\sum_{q\neq 1}\frac{1}{q-1}\left(\frac{\ell_0}{r_0}\right)^{(q-1)\varepsilon}\!\int\!\dd \Omega_{d-1}\,\varphi^{(\varepsilon)}_{3,q}(\mathbf{n})+ \mathcal{O}\left(\varepsilon\right)\,.
\end{equation}

\subsubsection{Far zone expansion of the potentials in $d$ dimensions}

From~\eqref{eq:DV_compact}, we see that the difference in regularization schemes depends on specific orders in the far-zone expansion of the integrand of the volume terms. As those are made of products of (derivatives of) potentials (defined in App. A of~\cite{MHLMFB20}), we need the expansion of individual potentials in $d$ dimensions.

The expressions of the simplest compact support potentials $V$, $V_i$, $K$ and the super-potentials (defined in Sec.~III.B of~\cite{MHLMFB20}) in the whole $d$-dimensional space are already available, so that we have just to expand them when $r \rightarrow +\infty$. However, the non-compact support potential $\hat{W}_{ij}$ is also required at 1PN order, and $\hat{X}$, $\hat{R}_i$ and $\hat{Z}_{ij}$ at Newtonian order, none of those potentials being known in all $d$-dimensional space. We thus have to compute their asymptotic behaviours by iterating the propagator at infinity and, crucially, add an appropriate homogeneous solution.

The far-zone expansion (or multipolar expansion, indicated by the operator $\mathcal{M}$) of a potential $P$ with source $S$ (assumed to be PN expanded), \textit{i.e.}, such that $\Box P=S=\overline{S}$ in $d$ dimensions, reads~\cite{BBBFMc}
\begin{equation}\label{eq:pot_matching}
\mathcal{M}(P) = \PF\,\Box^{-1}\left[\left(\frac{r}{r_0}\right)^B \mathcal{M}(S)\right] - \frac{1}{4\pi}\sum_{\ell\in\mathbb{N}}\frac{(-)^\ell}{\ell !}\partial_LS_\star^L\,.
\end{equation}
In principle, the first term is built from the PN-expanded retarded propagator $\Box_\text{R}^{-1}$, but we trade it here for the symmetric one $\Box^{-1}=\Box_\text{S}^{-1}$, since we are merely interested in even orders. 
The second term of the RHS is a homogeneous solution constructed out of the PN expansion of 
\begin{equation}
S_\star^L(t,r) \equiv \frac{\tilde{k}}{2r^{d-2}}\int_1^{+\infty}\!\dd z\,\gamma_\frac{1-d}{2}(z)\Bigl[\mathcal{S}_L\left(t-zr/c\right)+\mathcal{S}_L\left(t+zr/c\right)\Bigr]\,,
\end{equation}
where we recall that $\tilde{k}\equiv\pi^{1-\frac{d}{2}} \Gamma(\frac{d}{2}-1)$, and that $\gamma_\frac{1-d}{2}(z)$ is the kernel function entering the Green function of the d'Alembertian equation in $d$ dimensions, defined by Eq.~(3.3) in~\cite{BBBFMc} and recalled in Eq.~(2.3) of the follow-up paper~\cite{DDR_radiatif}. In other words, $S_\star^L(t,r)$ is an elementary monopolar homogeneous solution of the d'Alembertian equation, parametrized by the moments
\begin{equation}\label{eq:pot_SL}
\mathcal{S}_L(u) \equiv \PF\int\!\dd^d\mathbf{x}\left(\frac{r}{r_0}\right)^B x_L\,S(\mathbf{x},u)\,.
\end{equation}
Note that these particular moments are chosen to be non-STF, with just $x_L\equiv x_{i_1}\cdots x_{i_\ell}$. Performing explicitly the PN expansion, Eq.~\eqref{eq:pot_matching} becomes
\begin{align}\label{eq:matching_PN_exp}
\mathcal{M}(P) &= 
\sum_{k\in\mathbb{N}}\left(\frac{1}{c}\frac{\partial}{\partial t}\right)^{2k}\PF \Delta^{-k-1}\left[\left(\frac{r}{r_0}\right)^B \mathcal{M}(S)\right]\nonumber\\&\qquad - \frac{1}{4\pi^\frac{d-1}{2}}\sum_{k,\ell}\frac{(-)^\ell}{\ell!(2k)!}\frac{\Gamma\left(\frac{d}{2}-1-k\right)}{\Gamma\left(\frac{1}{2}-k\right)}\frac{1}{c^{2k}}\,\mathcal{S}_L^{(2k)}(t)\,\partial_Lr^{2k+2-d}\,, 
\end{align}
plus odd powers of $1/c$ which can be ignored for the present purpose.

Computing each of those terms requires different techniques:
\begin{itemize}
\item \textit{Particular solutions.} There are no specific issues with the multipolar expansion of the sources; so, computing the first term in~\eqref{eq:matching_PN_exp} is straightforwardly done by means of iterations of the $d$-dimensional ``\textit{Matthieu}'' formula~\cite{BDE04}
\begin{equation}\label{eq:Matthieu}
\Delta^{-1}\Bigl[ r^\alpha\, \hat{n}_{L}\Bigr] = \frac{r^{\alpha+2} \,\hat{n}_{L}}{(\alpha +d +\ell)(\alpha+ 2 -\ell)}\,.
\end{equation}
\item \textit{Homogeneous solutions.} The delicate part in the computation of expanded $d$-dimensional potentials is the control of the homogeneous solutions.
Indeed, the source integral $\mathcal{S}_L(u)$ of~\eqref{eq:pot_SL}, of non-compact support, is composed of terms like 
\begin{equation}\label{eq:pot_SL_int}
\int\!\dd^d\mathbf{x}\,x^L\,r^B\,r_1^\alpha\,r_2^\beta \, n_1^P n_2^Q\,.
\end{equation}
Those can be computed either by means of the generalized Riesz formulae described in Appendix~\ref{app:riesz} below, or by implementing the nice method of App. A in~\cite{HaSS13}, relying on the use of prolate spheroidal coordinates. In addition to the integrals~\eqref{eq:pot_SL_int}, we had to deal with the cubic sector of the potential $\hat{X}$, which is defined by Eq.~(A.4f) in~\cite{MHLMFB20} and whose source contains the delicate term
\begin{equation}
W_{ij}\, \partial_{ij}V \,.
\end{equation}
This cubic part requires \textit{a priori} the knowledge of the potential $\hat{W}_{ij}$ all over the space in $d$ dimensions at Newtonian order, but this has not been calculated yet. We could have done it relying on the generalization of the Fock function $g=\ln (r_1+r_2+r_{12})$ in $d$ dimensions, which has been derived in~\cite{BDE04}. However, its expression is only given in an integral form, which is not very convenient in practice. Instead, we employed the method of super-potentials~\cite{MHLMFB20} in order to replace $\hat{W}_{ij}$ by the expression of its source, which is now, indeed, known all over the space. Therefore, the asymptotic behaviours of required potentials have been fully determined with the appropriate accuracy.
\end{itemize}

Nevertheless, these computations are heavy\footnote{For instance, it took more than $160$ CPU hours to compute the 1PN $\hat{W}_{ij}$ up to $\mathcal{O}(r^{-4})$. At this order, its expression contains more than $10\,000$ terms, of which $\sim 8\,000$ come from the homogeneous part.} and some consistency checks are required. The first and most stringent one is that they were performed in a ``double-blind'' fashion. In addition, we investigated the three-dimensional limits of our potentials, confirming that they agree with the asymptotic expressions that were used to compute the three-dimensional surface terms in~\cite{MHLMFB20}. We have also verified that the harmonicity relations $\partial_\mu h^{\mu\nu} = 0$ hold, up to 1PN order and to the highest achievable order in $1/r$. At the 1PN order, those conditions read~\cite{MHLMFB20}
\begin{subequations}
	\begin{align}
	\partial_\mu \overline{h}^{\mu 0}=
	& \frac{d -  1}{2 (d -  2)}  \partial_t V +\partial_iV_i \\
	& + \frac{1}{c^2}\left[
	\partial_t\left(-\frac{(d -  1) (d -  3)}{(d-  2)^2} \,\hat{K} + \frac{\hat{W}}{2} + \frac{(d - 1)^2}{2 (d -  2)^2}\, V^2\right)
	+ \partial_i\left(2\hat{R}_i+\frac{d-1}{d-2}\,VV_i\right)\right]\,,\nonumber\\
	\partial_\mu \overline{h}^{\mu i}=
	& \partial_t V_{i} + \partial_j\left(\hat{W}_{ij}-\frac{\hat{W}}{2}\,\delta_{ij}\right)
	+ \frac{1}{c^2}\left[\partial_t\left(2 \hat{R}_{i} + \frac{d -  1}{d - 2} \,V V_{i}\right) 
	+ 4\,\partial_j\left(\hat{Z}_{ij}-\frac{\hat{Z}}{2}\,\delta_{ij}\right)\right]\,.
	\end{align}
\end{subequations}
The cancellation of $\partial_\mu h^{\mu 0}$ has been checked exactly at Newtonian order, and up to $\mathcal{O}(r^{-6})$ at 1PN order. The fact that $\partial_\mu h^{\mu i}$ vanishes has been checked up to $\mathcal{O}(r^{-7})$ at Newtonian order, and to $\mathcal{O}(r^{-5})$ at 1PN order.

\subsection{Compact terms}

An interesting and non-trivial feature of the IR regularization scheme is that it affects the evaluation of the potentials at the location of the particles. This is due to the fact that most of the potentials have a non-compact support source: the non-linear source terms extend towards infinity, so that the value of the potential at, say, $\bm{y}_1$ is sensitive to the IR regularization process. (Naturally, this effect does not impact the compact-support potentials, whose sources are proportional to Dirac distributions, and which are thus sensitive to the UV regularization only.) Therefore, the IR regularization scheme affects the compact terms that involve some non-compact potentials evaluated at $\bm{y}_1$ or $\bm{y}_2$.

Let us consider a potential $P$ with $d$-dimensional source $S$. As in the previous section, we are interested in the IR behaviour of $S$, hence we will expand it in the far zone. In practice, none of the sources we are interested in develop poles (or equivalently, logarithms in three dimensions); so, we can safely consider that
\begin{equation}\label{eq:S_comp}
S(\mathbf{x},t) = \sum_{p,q}\frac{\ell_0^{q\varepsilon}}{r^{p+q\varepsilon}}\,\varphi^{(\varepsilon)}_{p,q}(\mathbf{n},t)\,,
\end{equation}
where $\varphi^{(\varepsilon)}_{p,q}$ has no pole. The source takes the three-dimensional limit
\begin{equation}
S^{(d=3)}(\mathbf{x},t) = \sum_{p}\frac{\varphi_p(\mathbf{n},t)}{r^p}\,,
\quad \text{with} \quad
\varphi_p \equiv \sum_{q}\varphi^{(\varepsilon=0)}_{p,q}\,.
\end{equation}

Let us first deal with the Newtonian case, \textit{i.e.}, let us compute the difference induced by the change of IR regularization scheme in the Poisson integral evaluated in $\bm{y}_1$,
\begin{equation}\label{eq:Poisson_comp}
P(\bm{y}_1) = - \frac{\tilde{k}}{4\pi} \PF \int_{r'>\mathcal{R}}\!\dd^d\mathbf{x}'\left(\frac{r'}{r_0}\right)^B\frac{S\left(\mathbf{x}'\right)}{\vert\bm{y}_1-\mathbf{x}'\vert^{d-2}}\,,
\end{equation}
where we have safely replaced $\mathbf{x}$ by $\bm{y}_1$ in the kernel, as we integrate over the domain $r'\equiv\vert\mathbf{x}'\vert >\mathcal{R} \gg \vert\bm{y}_1\vert$, where we are free from UV divergences. Expanding the kernel at spatial infinity according to
\begin{equation}
\frac{1}{\vert\bm{y}_1-\mathbf{x}'\vert^{d-2}} = \sum_{\ell\in\mathbb{N}}\frac{2^\ell}{\ell!}\frac{\Gamma\left(\frac{d-2}{2}+\ell\right)}{\Gamma\left(\frac{d-2}{2}\right)}\,\frac{y_1^L\,\hat{n}_L}{{r'}^{d-2+\ell}}\,,
\end{equation}
and using the machinery developed for the treatment of volume terms, we find that the difference between the $B\varepsilon$ and Hadamard regularization schemes turns out to be 
\begin{equation}\label{eq:DP1_Newt}
\mathcal{D}P^\text{Newt}(\bm{y}_1) = 
- \sum_{\ell\in\mathbb{N}}\frac{(2\ell-1)!!}{4\pi \ell!}\,y_1^L\,\sum_{q\neq 0}\frac{1}{q}\left[\frac{1}{\varepsilon}-q \ln\left(\frac{r_0}{\ell_0}\right) + \sum_{k=0}^\ell	\frac{1}{2k-1}\right]\int\!\frac{\dd\Omega_{d-1}}{\Omega_{d-1}}\,\hat{n}_L\,\varphi^{(\varepsilon)}_{2-\ell,q}(\mathbf{n})\,,
\end{equation}
where we recall that the volume of the $(d - 1)$-dimensional sphere is $\Omega_{d-1}=2\pi^{d/2}/\Gamma(d/2)$. As expected, the scale $\mathcal{R}$ disappears from the difference in regularizations. The $q\neq 0$ criterion is a natural consequence of the derivation. The main structural difference with the formula for the volume terms~\eqref{eq:DV_compact} is the sum over $\ell$. Nevertheless, this sum is finite by virtue of the form of the source term~\eqref{eq:S_comp}, since $\ell$ is bounded by $2-p_0$.

The formula~\eqref{eq:DP1_Newt}, which is merely Newtonian, has to be generalized to higher PN orders. For this purpose, instead of starting from the Poisson integral~\eqref{eq:Poisson_comp}, we have to use the $d$-dimensional propagator. Again, we can restrict ourselves to an integration region $r'>\mathcal{R}\gg\vert\bm{y}_1\vert$ in which we are allowed to replace $\mathbf{x}$ by $\bm{y}_1$, so that
\begin{equation}
P(\bm{y}_1) = -\frac{\tilde{k}}{4\pi} \PF \int_{r'>\mathcal{R}}\!\frac{\dd^d\mathbf{x}'}{\vert\bm{y}_1-\mathbf{x}'\vert^{d-2}}\left(\frac{r'}{r_0}\right)^B
\!\int_1^{+\infty}\!\!\!\dd z\,\gamma_\frac{1-d}{2}(z)\,S\bigl(\mathbf{x}',t-z\vert\bm{y}_1-\mathbf{x}'\vert/c\bigr)\,,
\end{equation}
where the time-dependence of the source is now crucial. By PN-expanding this propagator, it can be seen that the action of the full propagator, in a formal PN sense, up to any PN order, is equivalent to the action of the mere Poisson integral~\eqref{eq:Poisson_comp} but acting on the effective source 
\begin{equation}
S^\text{eff}\left(\mathbf{x}',t,r_1'\right) = \int_1^{+\infty}\!\!\!\dd z\,\gamma_\frac{1-d}{2}(z)\,S\left(\mathbf{x}',t-z r_1'/c\right)\,,
\end{equation}
with $\mathbf{x}'$ just playing a spectator role. The PN expansion of this effective source is given by
\begin{subequations}\label{eq:Seff}
\begin{align}
&\overline{S}^\text{eff} \equiv \overline{S}^\text{eff}_\text{even} + \overline{S}^\text{eff}_\text{odd}\,,\\
& \overline{S}^\text{eff}_\text{even} = \sum_{j\in\mathbb{N}}\frac{\sqrt\pi}{(2j)!}\frac{\Gamma\left(\frac{1+\varepsilon}{2}-j\right)}{\Gamma\left(\frac{1+\varepsilon}{2}\right)\Gamma\left(\frac{1}{2}-j\right)}\left(\frac{r_1}{c}\right)^{2j}S^{(2j)}(\mathbf{x},t)\,,\label{eq:Seven}\\
& \overline{S}^\text{eff}_\text{odd} = \sum_{j\in\mathbb{N}}\frac{(-)^j}{j!}\frac{2\sqrt\pi\,\Gamma\left(\varepsilon\right)}{\Gamma\left(\frac{1+\varepsilon}{2}\right)\Gamma\left(-j-\frac{\varepsilon}{2}\right)\Gamma\left(2j+2+\varepsilon\right)}\left(\frac{r_1}{c}\right)^{1+2j+\varepsilon}\int_0^{+\infty}\!\dd\tau\,\tau^{-\varepsilon}S^{(2j+2)}(\mathbf{x},t-\tau)\,.
\end{align}
\end{subequations}
These expansion series correspond to the PN-even and PN-odd parts of the expansion of a homogeneous monopolar retarded solution of the wave equation in $d$ dimensions (see App.~A in~\cite{BBBFMc} for more details). As expected, the $j=0$ term of $S^\text{eff}_\text{even}$ is exactly given by $S$. Note that the PN-odd piece appears to be non-local in $d$ dimensions~\cite{BBBFMc}. Nevertheless, as already stated, we are merely interested in even terms\footnote{We will add the already known 2.5 and 3.5PN terms~\cite{FMBI12} to our final result, together with the 4PN dissipative non-local tail term (see Eqs. (6.11) and (6.12) in the companion paper~\cite{DDR_radiatif}).} and, thus, will not consider it. For a potential entering the source at a given PN order, we can simply apply the Newtonian result~\eqref{eq:DP1_Newt}, but using the effective source $S^\text{eff}_\text{even}$ truncated at the appropriate PN order following Eqs.~\eqref{eq:Seff}.

The required potentials that developp a non-vanishing difference at point $\bm{y}_1$ are $\hat{W}_1^{2\text{PN}}$, $\hat{Z}_1^{1\text{PN}}$, $\hat{X}_1^{1\text{PN}}$, $\hat{T}_1^\text{N}$ and $\hat{M}_1^\text{N}$ (with the index 1 denoting the value at particle 1 and the superscript the PN order) while the ones of the super-potentials vanish. The calculations reveal that all the non-vanishing corrections in our potentials are connected. Indeed, we found the interesting, but probably not very profound, relations
\begin{equation}
\mathcal{D}\hat{W}_1^{2\text{PN}}
=\frac{4}{c^2}\,\mathcal{D}\hat{Z}_1^{1\text{PN}}
=-\frac{2\varepsilon}{c^2}\,\mathcal{D}\hat{X}_1^{1\text{PN}}
=\frac{8}{3c^4}\,\mathcal{D}\hat{T}_1^\text{N}
=-\frac{16\varepsilon}{3c^4}\,\mathcal{D}\hat{M}_1^\text{N}\,.
\end{equation}
None of the other potentials receives corrections at the required order. These relations are valid up to the $\mathcal{O}(\varepsilon^0)$ order only, as the $\mathcal{O}(\varepsilon)$ remainders do not play any role in the IR dimensional regularization of the compact-support terms. Note that only the trace of the potential $\hat{M}_{ij}$ at Newtonian order (\textit{i.e.} $\hat{M}\equiv \hat{M}_{ii}$) as well as the potential $\hat{X}$ at 1PN order develop poles. Moreover, only the ``scalar'' sector of the potentials is affected: neither $\hat{R}_i$ nor $\hat{Y}_i$ are modified at the particles' positions; regarding the tensor potentials $\hat{Z}_{ij}$ and $\hat{M}_{ij}$, only their traces are impacted. In addition to those relations, the difference itself can be compactly written in terms of the Newtonian moment of inertia $I \equiv m_1 \bm{y}_1^2 + m_2 \bm{y}_2^2$ as
\begin{equation}
\mathcal{D}\hat{M}_1^\text{N} 
= \left(\frac{3}{8}-\varepsilon\right)c^2\,\mathcal{D}\hat{X}_1^{1\text{PN}}
= \frac{G^2(m_1+m_2)}{4}\left[\ln\left(\frac{r_0\sqrt{\bar{q}}}{\ell_0}\right)-\frac{1}{2\varepsilon}+\frac{1}{2}\right]\, I^{(4)}\,,
\end{equation}
where we recall that $\bar{q}\equiv 4\pi \de^{\gamma_E}$. The regularization induced differences of those potentials yield corrections in the effective mass $\tilde{\mu}_1$ (see Eq.~(2.17) in~\cite{MHLMFB20}), and in a few compact-support terms that all enter the source mass quadrupole moment at 4PN order.

\subsection{Surface terms}

Turning now to the surface terms, we will not calculate the difference between their values obtained in the two regularization schemes, but rather directly compute them in $d$ dimensions and show that they actually vanish. As presented in~\cite{MHLMFB20}, surface terms appearing in the mass quadrupole are of two kinds: ``Laplacian'' and ``divergence'' terms.
\begin{itemize}
\item A surface term of Laplacian type reads
\begin{equation}\label{eq:T}
T_L = \PF \intR \left(\frac{r}{r_0}\right)^B\,\hat{x}_L\,\Delta G\,,
\end{equation}
where $G$ is a $d$-dimensional function (in practice a product of potentials). As before, we have restricted ourselves to an integral over $r > \mathcal{R}$. Like for the volume terms, the function $G$ we will consider can be expanded near spatial infinity as:
\begin{equation}
G(\mathbf{x}) = \sum_{p,q}\frac{\ell_0^{q\varepsilon}}{r^{p+q\varepsilon}}\,\gamma^{(\varepsilon)}_{p,q}(\mathbf{n})\,,
\end{equation}
allowing the $\gamma^{(\varepsilon)}_{p,q}$ to contain poles. Inserting this expansion into the integral~\eqref{eq:T}, performing an integration by parts, using $\Delta(r^B \hat{x}_L) = B(B+d+2\ell-2)r^{B+\ell-2} \hat{n}_L$, and dropping the integrated terms that are vanishing by analytic continuation in either $B$ or $\varepsilon$ near zero, we are led to
\begin{equation}\label{eq:laplace}
T_L = -\left(2\ell+1+\varepsilon\right)\int\!\dd\Omega_{d-1}\hat{n}_L\,\gamma^{(\varepsilon)}_{\ell+1,1}(\mathbf{n})\,.
\end{equation}
\item A surface term of divergence type reads
\begin{equation}
K = \PF\intR\left(\frac{r}{r_0}\right)^B\,\partial_iH^i\,,
\end{equation}
where $H^i$ is a product of potentials or super-potentials. We again have the far-zone expansion
\begin{equation}
H^i(\mathbf{x}) = \sum_{p,q}\frac{\ell_0^{q\varepsilon}}{r^{p+q\varepsilon}}\,\eta^{(\varepsilon)i}_{p,q}(\mathbf{n})\,,
\end{equation}
where the $\eta^{(\varepsilon)i}_{p,q}$ can contain poles. A similar procedure yields
\begin{equation}\label{eq:divergence}
K = \int\!\dd\Omega_{d-1}n_i\,\eta^{(\varepsilon)i}_{2,1}(\mathbf{n})\,.
\end{equation}
\end{itemize}

The noteworthy point is that the $d$-dimensional values of both types of surface terms, in Eqs.~\eqref{eq:laplace} and~\eqref{eq:divergence}, are non zero only in the special case $q=1$. Now, as explicit in their expressions displayed in App.~C of~\cite{MHLMFB20}, the integrands of those terms are made of non-linear products of potentials. As such, the coefficients of their asymptotic expansions all bear $q \geqslant 2$, which make them vanish. 

More precisely, it is immediate to see, from the Green function $\Delta^{-1}\delta^{(d)}(\mathbf{x}-\bm{y}_1) = -\tilde{k}\,r_1^{-1-\varepsilon}/(4\pi)$, that the expansions of compact support potentials have $q=1$. Now, the sources of non-compact support potentials are made of products of compact support ones, and the iteration of the Poisson integrals cannot reduce the number of $q$'s. The latter fact can be understood from the Matthieu formula~\eqref{eq:Matthieu}, together with the fact that the expansion of the homogeneous solution bears $q=1$, as explicitly shown by Eq.~\eqref{eq:matching_PN_exp} where the second term is $\propto\partial_Lr^{2k+2-d}\propto r^{2k-\ell-1-\varepsilon}$. Thus, all individual potentials $V$, $V_i$, $\hat{W}_{ij}$, \textit{etc.} admit asymptotic expansions with $q \geqslant 1$. A similar argument applies to the case of super-potentials, which implies that all non-linear products of (derivatives of) potentials or super-potentials have $q \geqslant 2$.
	
The conclusion is that the surface terms are vanishing in $d$ dimensions, so that we have simply to subtract their Hadamard values from the final result:
\begin{equation}
\mathcal{D}I_{ij}^\text{Surf} = -I_{ij}^\text{Surf,Had}\,.
\end{equation}

\subsection{The ``extra'' term}

The argument developed in~\cite{MHLMFB20} to discard the contribution of the $B(d-3)$ piece of $I_{ij}$ [the last term in~\eqref{eq:Iij_d_dim}] does not \textit{a priori} hold with IR dimensional regularization; further investigation is required. We can restrict ourselves to an integration in the far zone $r > \mathcal{R}$, and recast
\begin{equation}\label{eq:Bd-3}
\hspace{-0.2cm} I_{ij}^\text{extra} = C_d\,\PF\sum_{k \in\mathbb{N}}\frac{1}{2^k\,k!}\frac{\Gamma\left(\frac{d}{2}+3\right)}{\Gamma\left(\frac{d}{2}+3+k\right)}\left(\frac{1}{c}\frac{\dd}{\dd t}\right)^{2k} \intR\left(\frac{r}{r_0}\right)^B\,B\,
\frac{\hat{x}_{aij} \,x_b}{r^{2-2k}} \,\overline{\Sigma}_{ab}\,,
\end{equation}
with $C_d=- \frac{2}{c^2}\frac{(d-3)(d+2)}{d(d-2)(d+4)}$. The source density $\overline{\Sigma}_{ab}$~\eqref{eq:tau}--\eqref{eq:Sigmad} is composed of two pieces: The first one, coming from the stress-energy tensor $T^{\mu\nu}$, involves Dirac distributions and, thus, does not enter our computation, since we take $r > \mathcal{R}$. The second piece, entailing the non-linear (NL) gravitational interactions $\Lambda^{ij}$, is composed, as such, of products of potentials. We expand it as
\begin{equation}\label{eq:NL}
\overline{\Sigma}_{ab}^\text{NL}(\mathbf{x},t) = \sum_{p,q}\frac{\ell_0^{q\varepsilon}}{r^{p+q\varepsilon}}\,\sigma_{p,q}^{(\varepsilon)ab}(\mathbf{n},t)\,,
\end{equation}
where the functions $\sigma_{p,q}^{(\varepsilon)ab}$ contain many PN orders and can develop poles. Inserting~\eqref{eq:NL} into~\eqref{eq:Bd-3}, we can perform the radial integration, which yields
\begin{align}
I_{ij}^\text{extra} = -C_d\sum_{k \in\mathbb{N}}\frac{1}{2^k\,k!}\frac{\Gamma\left(\frac{d}{2}+3\right)}{\Gamma\left(\frac{d}{2}+3+k\right)}\left(\frac{1}{c}\frac{\dd}{\dd t}\right)^{2k}\left[\int\!\dd\Omega_{d-1}\,\hat{n}_{aij} n_b \,\sigma_{5+2k,1}^{(\varepsilon)ab}(\mathbf{n},t)\right]\,.
\end{align}
Exactly as in the case of the surface terms, only the terms with $q=1$ contribute. However, recalling that $\sigma_{p,q}^{(\varepsilon)ab}$ is made of products of potentials, it cannot involve $q=1$ terms. We see therefore that, even with the present IR $B\varepsilon$ regularization scheme, the ``extra'' piece does not contribute:
\begin{equation}
 \mathcal{D}I_{ij}^\text{extra} = 0\,.
\end{equation}

\section{Final expression of the source quadrupole moment}\label{sec:final}

Summing up the pieces computed in the previous sections, and adding the local IR shift $\chi^i_{1,2}$ coming from the cancellation of the remaining poles in the conservative equations of motion~\cite{BBBFMc} (as described in Appendix~\ref{app:IR_shift}), we obtain the full dimensionally regularized source mass quadrupole moment at 4PN order,
\begin{equation}
I_{ij} = I_{ij}^\text{Had} + I_{ij}^\text{non-loc} + \mathcal{D}I_{ij} +\delta_\chi I_{ij}\,.
\end{equation}
Here, $I_{ij}^\text{Had}$ stands for the end result of~\cite{MHLMFB20},\footnote{Note that, after publishing our partial result in~\cite{MHLMFB20}, we have spotted an error in the $d$-dimensional computation of the value of the potential $\hat{R}_i$ at 1PN order evaluated at $\bm{y}_A$. This error induces a small change in the value of $I_{ij}^\text{Had}$ that has been taken into account in the final result of the renormalized mass quadrupole, which will be displayed in the companion paper~\cite{DDR_radiatif}.} then $I_{ij}^\text{non-loc}$ is the non-local part computed in Sec.~\ref{sec:nonlocal} and given by Eq.~\eqref{Itail}, $\mathcal{D}I_{ij}$ is the sum of the four contributions~\eqref{eq:DIij_def} whose calculation has been detailed above, and $\delta_\chi I_{ij}$ is the contribution of the IR shift of Appendix~\ref{app:IR_shift}. Remember that the non-local sector $I_{ij}^\text{non-local}$ is composed of both a direct tail contribution~\eqref{eq:Itaildirect} and a shift contribution [see Eq.~\eqref{eq:Itailshift}].

As the complete expression for the regularized source moment is long and not very enlightening nor interesting \textit{per se}, we do not display it, but rather discuss its most striking feature: the remaining poles. Indeed, the source moment $I_{ij}$ contains poles, which will have to be compensated by the proper treatment  in dimensional regularization of the non-linear interactions (such as tails-of-tails) relating the source and radiative moments. This will be proven in the companion paper~\cite{DDR_radiatif}.

While the non-local tail $I_{ij}^\text{non-loc}$ and shifted pieces $\delta_\chi I_{ij}$ are purely 4PN contributions, the effect of the dimensional regularization $\mathcal{D}I_{ij}$ starts already at the 3PN order. It can be written in a convenient form as
\begin{equation}\label{eq:poles_3PN}
\mathcal{D}I_{ij}^\text{3PN} = 
- \beta_I\,\frac{G^2M^2}{c^6} \left[\Pi_\varepsilon+\frac{246\,299}{44\,940}\right]\,I_{ij}^{(2)}
+ 2\beta_I\,\frac{G^2M}{c^6} \left[\Pi_\varepsilon+\frac{252\,599}{44\,940}\right]\,P_{\langle i}P_{j\rangle}\,,
\end{equation}
where we have used a notation similar to~\eqref{eq:notL} for the ``dressed'' pole:
\begin{equation}\label{eq:notPi}
\Pi_\varepsilon \equiv \ln\left(\frac{\sqrt{\bar{q}}\,r_0}{\ell_0}
\right) - \frac{1}{2\varepsilon}\,,\qquad\bar{q} \equiv 4\pi \de^{\gamma_E}\,.
\end{equation}
In Eq.~\eqref{eq:poles_3PN}, $P_i$ is the constant linear momentum and $\beta_I = -\frac{214}{105}$ is the coefficient associated with the renormalization of the mass quadrupole moment~\cite{GRoss10}. The appearance of this coefficient, which is known to be associated with the ``tail-of-tail'' interaction~\cite{B98tail}, indicates the soundness of the removal of the poles by the correct treatment in dimensional regularization of the non-linear interactions in the radiative quadrupole.

As for the poles showing up at the 4PN order, once taken the 1PN correction of Eq.~\eqref{eq:poles_3PN} into account, they can be expressed in the center-of-mass (CoM) frame as
\begin{equation}
I_{ij}^\text{pole,CoM,4PN} =  
-\frac{G^2M}{2c^8\varepsilon}\left[
\frac{12}{7}\,I_{a\langle i}^{(2)}\,I_{j\rangle a}^{(2)}
-\frac{24}{7}\,I_{a\langle i}^{(1)}\,I_{j\rangle a}^{(3)}-\frac{4}{7}\,I_{a\langle i}\,I_{j\rangle a}^{(4)}
+\frac{4}{3}I_{a\langle i}^{(3)}J_{j\rangle \vert a}\right]\,,
\end{equation}
where $J_{i \vert j}$ is the $d$-dimensional constant angular momentum defined in~\cite{HFB_courant}. Once again, the fact that this pole can be recast as a combination of non-linear interactions (in particular the expected coupling $M \times I_{ij} \times I_{ij}$ corresponding to ``tails-of-memory'') is a strong indication that it will be compensated by the proper dimensional-regularization treatment of the tail/memory effects. This will be the topic of the next paper~\cite{DDR_radiatif}.

\acknowledgments

We thank Gabriel Luz Almeida, Laura Bernard, Stefano Foffa, Sylvain Marsat, Rafael Porto and Riccardo Sturani for interesting discussions. We are grateful to Laura Bernard for providing us with the IR shift coming from the 4PN equations of motion (see Appendix~\ref{app:IR_shift}). F.L. received funding from the European Research Council (ERC) under the European Union’s Horizon 2020 research and innovation programme (grant agreement No 817791).

\appendix

\section{Generalized Riesz formulae}
\label{app:riesz}

The standard Riesz formula in arbitrary dimension $d$ (with $a, b \in \mathbb{C}$) reads:
\begin{align}
\int \dd^d x\, r_1^a r_2^b = \pi^{d/2} \frac{\Gamma(-\frac{a+b+d}{2})\Gamma(\frac{a+d}{2}) \Gamma(\frac{b+d}{2})}{\Gamma(-\frac{a}{2}) \Gamma(-\frac{b}{2})\Gamma(\frac{a+b}{2}+d)} r_{12}^{a+b+d}\,.\label{rieszstandard}
\end{align}
It is a priori valid when the integral converges, \textit{i.e.}, for $a+d>0$, $b+d>0$ and $a+b+d<0$. However, the result can be extended by analytic continuation everywhere but, possibly, a countable set of parameter values. On the other hand, this formula may be generalized to similar integrals involving additional STF angular factors $\hat{x}^L$ and various $r^2$. Among the most useful ones, we have
\begin{subequations}
\begin{align}
& \int \dd^d x\, \hat{x}^L r_1^a r_2^b = \pi^{d/2}
\frac{\Gamma(-\frac{a+b+d}{2})}{\Gamma(-\frac{a}{2}) \Gamma(-\frac{b}{2})}
\,r_{12}^{a+b+d} \sum_{s=0}^{\ell} \genfrac{(}{)}{0pt}{}{\ell}{s}
\frac{\Gamma(\frac{a+d}{2}+s) \Gamma(\frac{b+d}{2}+\ell-s)}
{\Gamma(\frac{a+b}{2}+d+\ell)} y_1^{\langle L-S}y_2^{S\rangle}\,,\label{rieszgeneral}\\
& \int \dd^d x\, r^2 \hat{x}^L r_1^a r_2^b = -\pi^{d/2} \frac{\Gamma(-\frac{a+b+d}{2}-1)}{\Gamma(-\frac{a}{2}) \Gamma(-\frac{b}{2})} \,r_{12}^{a+b+d} \sum_{s=0}^{\ell} \genfrac{(}{)}{0pt}{}{\ell}{s} \frac{\Gamma(\frac{a+d}{2}+s) \Gamma(\frac{b+d}{2}+\ell-s)} {\Gamma(\frac{a+b}{2}+d+\ell+1)} \times \nonumber \\ & \qquad\qquad ~ \times y_1^{\langle L-S}y_2^{S\rangle} \Big[\Big(\frac{b}{2}-s+1\Big) \Big(\frac{b+d}{2}+\ell-s\Big) y_1^2 + \Big(\frac{a}{2}-\ell+s+1\Big) \Big(\frac{a+d}{2}+s\Big) y_2^2 \nonumber\\& \qquad\qquad\qquad\qquad\qquad\qquad\qquad+ 2 \Big( \frac{a+d}{2} +s \Big) \Big( \frac{b+d}{2} +\ell-s \Big) (y_1y_2) \Big] \,,
\end{align}
\end{subequations}
with $\genfrac{(}{)}{0pt}{}{\ell}{s}$ denoting the usual binomial coefficient, and $y_A^2=\bm{y}_A^2$, $(y_1y_2)=\bm{y}_1\cdot\bm{y}_2$.

Let us sketch for instance the proof of the first one. The starting point consists in rewriting $x^L$ as $(r_1 n_1 + y_1)^L \equiv (r_1 n_1^{i_1} + y_1^{i_1})\cdots (r_1 n_1^{i_\ell} + y_1^{i_\ell})$, where $n^i_1$ is the unit vector $n^i_1=(x^i-y_1^i)/r_1$, and expanding the product. After permuting the integration and summation symbols, we get a sum of elementary integrals of the form
\begin{equation}
\int  \dd^d x\, r_1^{a+s} r_2^b \, \hat{n}_1^L \, , 
\end{equation}
for a summation index $0\leqslant s \leqslant \ell$, with $n_1^L\equiv n_1^{i_1}\cdots n_1^{i_\ell}$. For each integral, the parameters $a$, $b$ and $d$ are chosen in a domain of $\mathbb{C}^3$ where the convergence is guaranteed. We can always manage to avoid the situation where one of them is an integer. We then express the factors $r_1^{a+s}\hat{n}_1^L$ as a multi-derivative using the relation
\begin{align} \label{eq:derr1}
\hat{\partial}_L r_1^\alpha = (-)^\ell \hat{\partial}_{1L} r_1^\alpha = 2^\ell \frac{\Gamma(\alpha/2+1)}{\Gamma(\alpha/2-\ell+1)} r_1^{\alpha-\ell} \hat{n}_1^L \, ,
\end{align}
with $\partial_{1i} = \partial/\partial y_1^i$, valid as long as $\hat{\partial}_{1L} r_1^\alpha$ does not involve any distributional contribution, which is indeed the case when $\alpha$ is a non-integer. Next, we commute the derivatives and the integral (taking for instance $2s<d$, $2s<-a<d$ and $a+d<-b<d$ to bypass any convergence issue). We find
\begin{align} 
\int \dd^d x\, \hat{x}^L r_1^a r_2^b = \sum_{s=0}^{\ell} \genfrac{(}{)}{0pt}{}{\ell}{s} \frac{(-)^s}{2^s} \frac{\Gamma(a/2+1)}{\Gamma (a/2+s+1)} y_1^{\langle L-S} \partial_{1S\rangle}  \int \dd^d x\,  r_1^{a+2s} r_2^b\, . 
\end{align}
We now use the standard Riesz formula~\eqref{rieszstandard}, which produces a dimensional factor $r_{12}^{a+b+d+2s}$, and apply the derivative $\hat{\partial}_{1S}$ as in Eq.~\eqref{eq:derr1}. At this stage, we simplify some factors under the sum by noticing that, for $s \in \mathbb{N}$,
\begin{align}
\Gamma(z+s+1) \Gamma(-z-s) = \frac{\pi}{\sin[- (z +s)\pi]} = (-)^s \Gamma(z+1) \Gamma(-z) \, ,
\end{align}
and expand the position-dependent factor $r_{12}^{a+b+d+s} n_{12}^{\langle S}\, y_1^{L-S\rangle} = r_{12}^{a+b+d} (y_1-y_2)^{\langle S} y_1^{L-S\rangle}$. The result is obtained by resummation:\footnote{It is a mere consequence of the relation ${}_2F_1 (a,b;c;1) = \frac{\Gamma(c) \Gamma(c-a-b)}{\Gamma(c-a) \Gamma(c-b)}$ for $\Re(c-a-b)>0$, where ${}_2F_1 (a,b;c;z)$ is the Gaussian hypergeometric function, with standard notations.}
\begin{align}
& \sum_{s=0}^{+\infty} \frac{(-)^s}{(\ell-k-s)! s!} \frac{\Gamma(\frac{a+d}{2}   +s+k)}{\Gamma(\frac{a+b}{2}+d+s+k)} \nonumber \\ & \qquad \qquad \qquad = \frac{\Gamma(\frac{a+d}{2} +k)}{(\ell-k)! \Gamma(\frac{a+b}{2}+d+k)} \frac{\Gamma(\frac{a+b}{2}+d+k) \Gamma(\frac{b+d}{2}+\ell-k)}{ \Gamma(\frac{a+b}{2}+d+\ell) \Gamma(\frac{b+d}{2})} \, ,
\end{align}
for $k$ non-negative integer lower than $\ell$.

A third particularly useful generalization of Eq.~\eqref{rieszstandard} reads
\begin{align} \label{rieszgeneral2}
& \int \dd^d x\, r_1^a \,r_2^b \,n_1^L n_{2\, K} = \pi^{d/2} \frac{\Gamma(-\frac{a+b+\ell+k+d}{2}) \Gamma(\frac{a+b+\ell+k+d}{2}+1)} {\Gamma(-\frac{a-\ell}{2}) \Gamma(-\frac{b-k}{2}) \Gamma(\frac{a+b+\ell+k}{2}+d)} r_{12}^{a+b+d} \times \nonumber \\ & \qquad\qquad ~ \times \sum_{i=0}^{[\frac{\ell+k}{2}]} \sum_{j=\max(0,2i-k)}^{\min{(2i,\ell)}} \sum_{p=\max(0,j-i)}^{[\frac{j}{2}]}\!\!\!\!\! \frac{\Gamma(\frac{a+\ell+d}{2}+i-j) \Gamma(\frac{b+k+d}{2}+j-i)} {\Gamma(\frac{a+b-\ell-k+d}{2}+i+1)}\times \nonumber\\&\qquad\qquad\qquad\qquad\qquad\qquad \times\frac{(-)^{k+i+j} \ell! k! \, \delta_{(2I2P-2J} n_{12\, KJ-2I} \delta_{J-2P)}^{(J-2P} n_{12}^{L-J} \delta^{2P)}} {2^{2i+2p-j} (\ell-j)!(k+j-2i)!p!(i+p-j)!(j-2p)!}\,,
\end{align}
with the convention that the contravariant indices belong to $\{L\}=\{i_1,\cdots,i_\ell\}$ and the covariant ones to $\{K\}=\{j_1,\cdots, j_k\}$, while $\delta^{J-2P}_{J-2P}= \delta^{a_1}_{b_1}\cdots \delta^{a_{j-2p}}_{b_{j-2p}}$, $\delta^{2P} = \delta^{a_{j-2p+1} a_{j-2p+2}} \cdots \delta^{a_{j-1}a_j}$, and $\delta_{2I2P-2J}  =\delta_{b_{j-2p+1} b_{j-2p+2}}\cdots \delta_{b_{2i-j-1}b_{2i-j}}$. The upper (lower) symmetrization only affects the contravariant (covariant) indices, respectively.

The most delicate task in the derivation is the symmetrization over the multi-indices $L$ and $K$ of the multi-derivative
\begin{align}
\delta^{(2S} \partial_{1}^{L-2S)} \delta_{(2R} \partial_{2\, K-2R)}
r_{12}^{a+b+\ell+k-2(s+r)+d}\, .
\end{align}
It is achieved with the help of the convenient identity
\begin{align}
\delta^{(2I} n_{12}^{L-2S} n_{12\, K-2R-2I)} &= i!
\frac{(\ell + k-2s-2r-2i)!}{(\ell+k-2s-2r)!} \sum_{j=\max(0,2i+2r-k)}^{\min (2i,\ell-2s)}
\sum_{p=\max(0,j-i)}^{[\frac{j}{2}]} \frac{1}{2^{2p-j}p!} \times \nonumber \\
& \qquad  \times \frac{(\ell-2s)!}{(\ell-2s-j)!} \frac{(k-2r)!}{(k+j-2r-2i)!}
\frac{1}{(i+p-j)!(j-2p)!} \times \nonumber \\ & \qquad \times \delta_{(2P2I2R-2J} n_{12\, KJ-2R-2I} \delta_{J-2P)}^{(J-2P} n_{12}^{L-2S-J}\delta^{2P)} \, .
\end{align}
In this manner, we find that the integral on the left-hand side of Eq.~\eqref{rieszgeneral2}, after the appropriate change of indices, may be put in the form
\begin{align}
\sum_{i=0}^{[\frac{\ell+k}{2}]} \sum_{j=\max(0,2i-k)}^{\min(2i,\ell)} \sum_{p=\max(0,j-i)}^{[\frac{j}{2}]}
\sum_{s=0}^p \sum_{r=0}^{p+i-j} r_{12}^{a+b+d} \,\lambda_{s,r}^{i,j,p} \,\delta_{(2I2P-2J} n_{12\, KJ - 2I} \delta_{J-2P)}^{(J-2P} n_{12}^{L-J} \delta^{2P)}\, ,
\end{align}
with
\begin{align}
\lambda_{s,r}^{i,j,p} &=  \frac{(-)^{s+r+\ell}2^{j-2p-2i}}{(\ell -j)!(k+j-2i)! (p-s)! (i-r+p-j)! (j-2p)!} \frac{\ell! k!}{s! r!} \frac{\Gamma(\frac{a-\ell}{2}+1)}{\Gamma(\frac{a+\ell}{2}+1-s)}  \times \nonumber \\ & \times \frac{\Gamma(\frac{b-k}{2}+1)}{\Gamma(\frac{b+k}{2}+1-r)} \frac{\Gamma(\frac{a+b+\ell+k+d}{2}-s-r+1)}{\Gamma(\frac{a+b-\ell-k+d}{2}+i+1)}  \bigg[\int \dd^d x\, r_1^{a+\ell-2s} r_2^{b+k-2r}\bigg]_{r_{12}=1} \, .
\end{align}
Resorting to techniques similar to those employed to compute the previous integrals, we can perform explicitly the sums over $s$ and $r$,
\begin{align}
\sum_{s=0}^{+\infty} \sum_{r=0}^{+\infty}  \lambda_{s,r}^{i,j,p} &=  \frac{\ell! k!}{2^{2i+2p-j} (\ell-j)! (k+j-2i)!} \frac{\pi^{d/2} (-)^{k+i+j}}{p! (i+p-j)! (j-2p)!} \times \nonumber \\ & \times \frac{\Gamma(-\frac{a+b+\ell+k+d}{2}) \Gamma(\frac{a+b+\ell+k+d}{2}+1)} {\Gamma(-\frac{a-\ell}{2}) \Gamma(-\frac{b-k}{2}) \Gamma(\frac{a+b+\ell+k}{2}+d)} \frac{\Gamma(\frac{a+\ell+d}{2}+i-j) \Gamma(\frac{b+k+d}{2}+j-i)} {\Gamma(\frac{a+b-\ell-k+d}{2}+i+1)} \, ,
\end{align}
which yields the desired result.

\section{The local piece of the IR shift}
\label{app:IR_shift}

In addition to the non-local shift due to tails, as obtained in Eq.~\eqref{eq:nlshift}, the source quadrupole has to be shifted by the exact same IR shift that was applied to the conservative sector to compensate the remaining poles in the equations of motion~\cite{BBBFMc}. This IR shift $\chi^i_{1,2}$ is local and starts at the 4PN order. Thus, its contribution to the source quadrupole moment is simply given by $\delta_\chi I_{ij} = 2m_1\,y_1^{\langle i}\chi_1^{j\rangle} + 1\leftrightarrow 2$. The local IR shift $\chi_1^i$ can be decomposed as
\begin{equation}\label{eq:shift_local}
\chi_1^i = \frac{1}{c^8}\left(\frac{1}{\varepsilon}\,\chi_1^{i\,(-1)} +   \chi_1^{(0,y_{1})}\, y^i_{1}+ \chi_1^{(0,n_{12})}\, n^i_{12} +\chi_1^{(0,v_{1})}\,v^i_1 +\chi_1^{(0,v_{12})}\,v^i_{12}\right)\,.
\end{equation}
Recalling that $\bar{q} \equiv 4\pi e^{\gamma_E}$, $m \equiv m_1 +m_2$ and $\nu \equiv m_1m_2/m^2$, the various terms of Eq.~\eqref{eq:shift_local} are given by
\begin{subequations}
\begin{align}
\chi_1^{i\,(-1)} &= 
\frac{8 G^4 m^3\nu}{5 r_{12}{}^4} \bigg[m (n_{12} y_1)n_{12}^i+r_{12} (2 m_1-m_2)n_{12}^i
-\frac{m}{3}\,y_1^i\bigg]
\nonumber\\
&
-\frac{4 G^3 m^2\nu \,n_{12}^i}{5 r_{12}^3}\bigg[
2 v_{12}^2 \left(r_{12} (m_1-2 m_2)-3m \left(n_{12} y_1\right)\right)
-8 m r_{12} \left(v_1 v_{12}\right)\nonumber\\
& \hspace{2cm}
+\left(n_{12} v_{12}\right)^2 \left(30 m\left(n_{12} y_1\right) +9 m_2 r_{12}\right)
+6 m \left(n_{12} v_{12}\right) \left(2 r_{12} \left(n_{12} v_1\right)-3 \left(y_1 v_{12}\right)\right)\bigg]\nonumber\\
&
+ \frac{4G^3 m^2\nu\,v_{12}^i}{5 r_{12}{}^3}\bigg[8m \left(r_{12} \left(n_{12} v_1\right)-(y_1 v_{12})\right)+\left(n_{12} v_{12}\right) \left(18 m\left(n_{12} y_1\right)+5 m_2
r_{12}\right)\bigg]\nonumber\\   
& 
-\frac{8 G^3m^3\nu\,y_1^i}{15 r_{12}^3} \bigg[3 \left(n_{12} v_{12}\right)^2-v_{12}^2\bigg]
-\frac{16 G^3 m^3\nu \left(n_{12} v_{12}\right)}{15 r_{12}^2}\,v_1^i\,,\\
\chi_1^{(0,y_{1})} = &
\frac{G^4m^4\nu}{75 r_{12}^4} \left[80 \ln \left(\frac{r_0^2\bar{q}}{\ell_0^2}\right)+80 \ln\left(\frac{r_{12}}{r_0}\right)+73\right]\nonumber\\
& 
+ \frac{G^3m^3\nu}{75 r_{12}^3} \bigg[
\left(n_{12} v_{12}\right)^2 \left(180 \ln \left(\frac{r_0^2\bar{q}}{\ell_0^2}\right)+120 \ln \left(\frac{r_{12}}{r_0}\right)+359\right)\nonumber\\
&
\hspace{2cm} -v_{12}^2 \left(60 \ln\left(\frac{r_0^2\bar{q}}{\ell_0^2}\right)+40 \ln\left(\frac{r_{12}}{r_0}\right)+133\right)\bigg]\,,\\
\chi_1^{(0,n_{12})} = &
-\frac{G^4m^3\nu}{2100 r_{12}^4} 
\bigg[r_{12} \left(6720 (2 m_1-m_2)\left(\ln \left(\frac{r_0^2\bar{q}}{\ell_0^2}\right)+\ln \left(\frac{r_{12}}{r_0}\right)\right)+8861 m_1-5183 m_2\right)\nonumber\\
& \hspace{2cm}
+42m \left(n_{12} y_1\right)  \left(160 \ln \left(\frac{r_0^2\bar{q}}{\ell_0^2}\right)+160 \ln \left(\frac{r_{12}}{r_0}\right)+61\right)\bigg]\nonumber\\
&
+ \frac{G^3m^2\nu}{4200 r_{12}^3} \bigg[
2 r_{12}v_{12}^2  \left((m_1-2 m_2)\left(5040  \ln \left(\frac{r_0^2\bar{q}}{\ell_0^2}\right)+3360 \ln \left(\frac{r_{12}}{r_0}\right)\right)-121916 m_1-37995 m_2\right)\nonumber\\
& \hspace{2cm}
-84mv_{12}^2 \left(n_{12} y_1\right)  \left(360 \ln \left(\frac{r_0^2\bar{q}}{\ell_0^2}\right)+240\ln \left(\frac{r_{12}}{r_0}\right)+223\right)\nonumber\\
& \hspace{2cm}
+2100m\left(n_{12} v_{12}\right)^2\left(n_{12} y_1\right)  \left(72 \ln \left(\bar{q} r_0{}^2\right)+48 \ln\left(\frac{r_{12}}{r_0}\right)+35\right)\nonumber\\
& \hspace{2cm}
+3 r_{12}\left(n_{12} v_{12}\right)^2\left(15120 m_2 \ln \left(\frac{r_0^2\bar{q}}{\ell_0^2}\right)+10080 m_2 \ln \left(\frac{r_{12}}{r_0}\right)+126959 m_1+45908 m_2\right)\nonumber\\
& \hspace{2cm}
+21r_{12} \left(n_{12} v_1\right)\left(n_{12} v_{12}\right) (m_1+m_2)\left(2880 \ln \left(\frac{r_0^2\bar{q}}{\ell_0^2}\right)+1920 \ln \left(\frac{r_{12}}{r_0}\right)-9661\right)\nonumber\\
& \hspace{2cm}
-252 m \left(n_{12} v_{12}\right)\left(y_1 v_{12}\right)
\left(360 \ln\left(\frac{r_0^2\bar{q}}{\ell_0^2}\right)+240 \ln \left(\frac{r_{12}}{r_0}\right)+223\right)\nonumber\\
& \hspace{2cm}
-42m r_{12}   \left(v_1 v_{12}\right) \left(960 \ln \left(\frac{r_0^2\bar{q}}{\ell_0^2}\right)+640 \ln \left(\frac{r_{12}}{r_0}\right)-3007\right)\bigg]\,,\\
\chi_1^{(0,v_1)} = &
\frac{G^3 m^3\nu\left(n_{12} v_{12}\right)}{600 r_{12}^2} \left[960 \ln\left(\frac{r_0^2\bar{q}}{\ell_0^2}\right)+640 \ln\left(\frac{r_{12}}{r_0}\right)+2113\right]\,,\\
\chi_1^{(0,v_{12})} = &
-\frac{G^3 m^2\nu}{4200 r_{12}^3}\bigg[
5 r_{12}\left(n_{12} v_{12}\right) \left(5040 m_2 \ln\left(\frac{r_0^2\bar{q}}{\ell_0^2}\right)+3360 m_2 \ln\left(\frac{r_{12}}{r_0}\right)+55132 m_1+41681 m_2\right)\nonumber\\
&\hspace{2cm}
+ 42 m r_{12} \left(n_{12} v_1\right) \left(960 \ln\left(\frac{r_0^2\bar{q}}{\ell_0^2}\right)+640 \ln\left(\frac{r_{12}}{r_0}\right)-3007\right)\nonumber\\
&\hspace{2cm}
+ 252 m\left(n_{12} v_{12}\right)\left(n_{12} y_1\right)\left(360 \ln\left(\frac{r_0^2\bar{q}}{\ell_0^2}\right)+240 \ln  \left(\frac{r_{12}}{r_0}\right)+223\right)\nonumber\\
&\hspace{2cm}
-366 m \left(y_1 v_{12}\right) \left(120 \ln\left(\frac{r_0^2\bar{q}}{\ell_0^2}\right)+80 \ln\left(\frac{r_{12}}{r_0}\right)+101\right)\bigg]\,.
\end{align}
\end{subequations}

\bibliography{ListeRef_DiffIR.bib}

\end{document}